\documentclass[12pt,aps,eqsecnum,nofootinbib,floatfix]{revtex4}
\UseRawInputEncoding
\usepackage [latin1]{inputenc}
\usepackage{amssymb}
\usepackage{amsmath}
\usepackage{graphicx}
\usepackage{dcolumn}
\usepackage{bm}
\usepackage{subfigure}
\usepackage{color}

\begin{document}

\title{Phase transition and entropic force in Reissner-Nordstr\"om-de Sitter
spacetime }
\author{Yang Zhang}
\author{Yu-bo Ma}
\author{Yun-zhi Du}
\email{duyzh13@lzu.edu.cn}
\author{Huai-fan Li}
\email{huaifan999@163.com}
\author{Li-chun Zhang}
\affiliation{ Institute of Theoretical Physics, Shanxi Datong University, Datong,
037009, China}

\vspace{-1.5cm}

\begin{abstract}
In this paper, thermodynamic properties of the Reissner-Nordstr\"om-de Sitter (RN-dS) black hole have been studied on the basis of the correlation between the black hole and cosmological horizons. It is found that the RN-dS black hole experiences a phase transition, when its state parameters satisfy certain conditions. From the analysis of the interaction between two horizons in RN-dS spacetime, we get the numerical solution of the interaction between two horizons. It makes us to realize the force between the black hole and cosmological horizons, which can be regarded as a candidate to explain our accelerating expansion universe. That provides a new window to explore the physical mechanism of the cosmic accelerating expansion.
\end{abstract}

\maketitle


\section{Introduction}

Since the discovery of the accelerating expansion cosmic, our four-dimensional universe in the early period of inflation was
regarded as a quasi-dS spacetime. Moreover, the Hawking-Page phase transition between a stable large black hole and a pure thermal radiation AdS spacetime was explained as the confinement/decomfinement phase transition of a gauge field by AdS/CFT. That urges people to look for the similar dual relations in de Sitter spacetime. Therefore, it is not only for the theoretical interesting but also for the practical necessity to study the de Sitter spacetime.

With the development of science and technology, the scientists use satellites, lasers, large astronomical telescopes, and supercomputers to observe and research the universe. New discoveries constantly refresh our understanding of the universe. The answer for the cosmic accelerating expansion is still not clear (gravity can only slow down the cosmic expansion). At present, a mainstream opinion is that, dark energy may be the answer, whose distribution is homogeneous on cosmic scales and accounts for most part of the cosmic total energy. However, it is only a hypothesis. What is dark energy? Where does it come from? And How it accelerates the cosmic expansion. These are the question which are still not clear.

Recently, with the research of dark energy, the black hole thermodynamic in a de Sitter spacetime has attracted much more
attentions \cite{1,2,3,4,5,6,7,8,9,10,11,12,13,14,15,16,17,18,19,20,21,22,23,24,25,26,27,28}. By regarding the cosmological
constant as dark energy, our universe maybe evolve into a new de Sitter phase. In order to reconstruct a whole cosmic expansion history and explore the internal cause for the cosmic accelerating expansion, we should have to study de Sitter spacetime on classical level, quantum level, and thermodynamic aspect. Moreover, the effect of parameters on the evolution of the de Sitter spacetime also should be concerned.

In this paper, by studying the thermodynamic characteristics, it is found that the RN-de Sitter (RN-dS) spacetime have the
first-order and second-order phase transitions as well as the charged AdS black holes. The phase transition point of RN-dS spacetime is related with the electric potential at the dS black hole horizon. It indicates that the evolution processes of the RN-dS spacetime tending to a pure dS spacetime are different for different electric potentials. Therefore, the electric potential at the dS black hole horizon plays a key role in the evolution of RN-dS spacetime. That will provide a new way to simulating the
accelerating expansion cosmic through the evolution process for the RN-dS spacetime.

By exploring the thermodynamic behaviors of the RN-dS spacetime, it is found that there exists an interaction between the two horizons. When the cosmological constant $l^{2}$ is fixed, the interaction between two horizons varies with the position ratio of them. When the two horizons are close to each other, there is a strong repulsive force between them. However, when the position ratio of two horizons tends to zero, i.e., when it tends to be the pure dS spacetime, the force between the two horizons disappears. This discovery provides a new way for people to explore the internal factors of the cosmic expansion.

This paper is arranged as follows. In Sec. II, the effective thermodynamic quantities in RN-dS spacetime will be given. Relationships between the radiation temperatures on the black hole horizon $T_{+}$ and the cosmological horizon $T_{c}$, and the position ratio of the two horizons are studied. Moreover, the relationship between the effective temperature of the spacetime $T_{eff}$ and the position ratio of horizons are also studied. In Sec. III, we extend the study method of phase transition in Van de Waals system or charged AdS black hole to the RN-dS spacetime. In Sec. IV, studying the relationship between the entropic force arisen from the interaction between the two horizons and the position ratio of the two horizons. The effect of the value for the electric potential on the black hole horizon and spacetime parameters on the entropic force will also be explored. The discussion and conclusions will be presented in Sec. V. (we use the units $G=\hbar =k_{B}=c=1$)

\section{ Effective thermodynamic quantities in RN-dS spacetime}

For the four-dimensional RN black hole embedded in a dS spacetime, the static spherically symmetric solution was shown as \cite{8}
\begin{equation}
d{s^{2}}=-f(r)d{t^{2}}+{f^{-1}}d{r^{2}}+{r^{2}}d\Omega _{2}^{2}  \label{2.1}
\end{equation}
with
\begin{equation}
f(r)=1-\frac{{{2}M}}{{r}}+\frac{{{Q^{2}}}}{{{r^{2}}}}-\frac{{r^{2}}}{{l^{2}}},
\end{equation}%
where $M$ and $Q$ are the black hole mass and charge, $l$ is the curvature
radius of dS space. When $f({r_{c}})=0$ and $f({r_{+}})=0,$ we have
\begin{equation}
M = \frac{{{r_c}(1 + x)}}{4} + \frac{{{Q^2}(1 + x)}}{{4{r_c}x}} - \frac{{%
r_c^3}}{{4{l^2}}}(1 + {x^3}),
\end{equation}
\begin{equation}
\frac{1}{{l^2}} = \frac{1}{{r_c^2(1 + x + {x^2})}} - \frac{{Q^2}}{{r_c^4x(1
+ x + {x^2})}},
\end{equation}
where $r_+$ and $r_c$ are the locations of the black hole and cosmological horizons, respectively, and $x=\frac{r_+}{r_c}$.
Considering the interaction between the black hole and cosmological horizons, the effective thermodynamic quantities
and corresponding first law of black hole thermodynamics can be given as
\begin{equation}
dM = {T_{eff}}dS - {P_{eff}}dV + {\phi _{eff}}dQ.
\end{equation}
Here the mass parameter of black hole is identified with the enthalpy. The thermodynamic volume is the one between the black hole and cosmological horizons \cite{1,3,4,15,2,29}
\begin{equation}
V = \frac{{4\pi }}{3}\left( {r_c^3 - r_ + ^3} \right).
\end{equation}
Since the entropy of the black hole horizon and the cosmological horizon is
only an explicit function of the location ratio of them, we set the total
entropy of our considering system as
\begin{equation}
S = \pi r_c^2(1 + {x^2} + f_0(x)),
\end{equation}
where the first two terms on the right side of above equation represent the entropy on the two horizons, respectively, and the undefined function $f_0(x)$ represents the extra contribution from
the correlations of the two horizons.

Taking Eqs. (2.3), (2.6) and (2.7) into Eq. (2.5), the effective temperature
$T_{eff}$ and the effective pressure $P_{eff}$ are obtained as
\begin{eqnarray}
{T_{eff}}&=&\frac{{1-x}}{{4\pi {r_{c}}x(1+{x^{4}})}}\left[(1+x)(1+{x^{3}})-2{x^{2}} -\frac{{{Q^{2}}}}{{r_{c}^{2}{x^{2}}}}{((1+x+{x^{2}})(1+{x^{4}})-2{x^{3}})}\right]\\
{P_{eff}} &=&\frac{{(1-x)(1+{x^{2}}+f_0(x))}}{{8\pi r_{c}^{2}x(1+{x^{4}})(1+x+{x^{2}})}}{\left( {(1+2x)-\frac{{{Q^{2}}(1+2x+3{x^{2}})}}{{r_{c}^{2}{x^{2}}}}}%
\right) } \nonumber\\
&&-\frac{{(1-x^2)(x+f^{\prime }_0(x)/2)}}{{8\pi r_{c}^{2}x(1+{x^{4}})}}{\left( {x+\frac{{{Q^{2}}(1+{x^{2}})}}{{r_{c}^{2}x}}}\right)
},
\end{eqnarray}
where
\begin{equation}
f_0(x)=\frac{8}{5}{\left( {1-{x^{3}}}\right) ^{2/3}}-\frac{{2\left( {4-5{x^{3}}%
-{x^{5}}}\right) }}{{5\left( {1-{x^{3}}}\right) }}.
\end{equation}
For the detailed calculating process on the above quantities, please refer to Ref. \cite{22}. It is easy to see that the Smarr's formula of the above thermodynamical quantities becomes
\begin{eqnarray}
 M(S,V,Q)=2T_{eff}S-3VP_{eff}+Q\phi_{eff}.
\end{eqnarray}

Then the radiation temperatures on the black hole horizon and the cosmological horizon are
\begin{eqnarray}
{T_{+}}&=&\frac{{f^{\prime }({r_{+}})}}{{4\pi }}=\frac{{1-x}}{{4\pi {r_{c}}%
x(1+x+{x^{2}})}}\left( {1+2x-\frac{{{Q^{2}}(1+2x+3{x^{2}})}}{{r_{c}^{2}{x^{2}%
}}}}\right),\\
{T_{c}}&=&-\frac{{f^{\prime }({r_{c}})}}{{4\pi }}=\frac{{(1-x)}}{{4\pi {r_{c}}%
(1+x+{x^{2}})}}\left( {2+x-\frac{{{Q^{2}}}}{{r_{c}^{2}x}}(3+2x+{x^{2}})}%
\right).
\end{eqnarray}%
When they are equal to each other, we find that \cite{15}
\begin{equation}
T={T_{+}}={T_{c}}=\frac{{1-x}}{{2\pi {r_{c}}{{(1+x)}^{2}}}}.
\end{equation}%
Considering Eq. (2.4), we get
\begin{equation}
\frac{1}{{r_{c}^{2}}}=\frac{{(1+x+{x^{2}})}}{{{l^{2}}(1-x{Q^{2}}/r_{+}^{2})}}%
.
\end{equation}%
\begin{figure}[tbp]
\centering
\subfigure{{\includegraphics[width=0.3\columnwidth,height=1.5in]{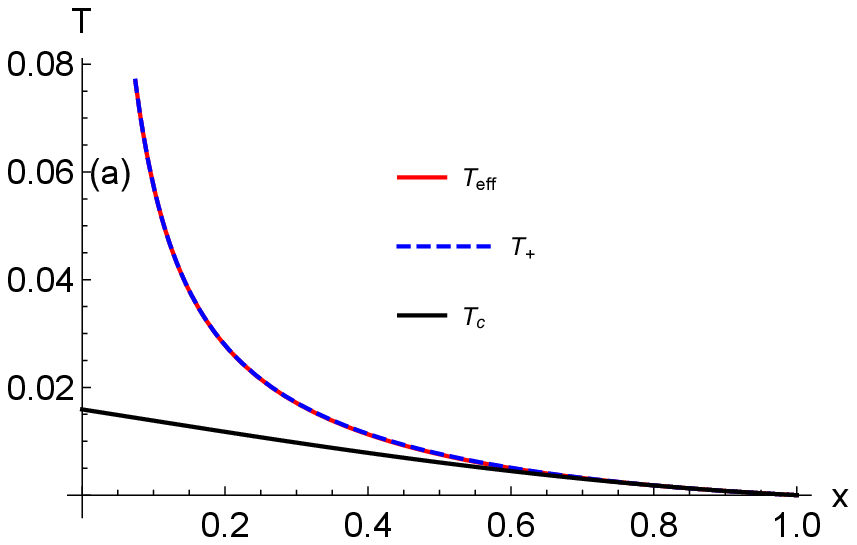}}}
\subfigure{{\includegraphics[width=0.3\columnwidth,height=1.5in]{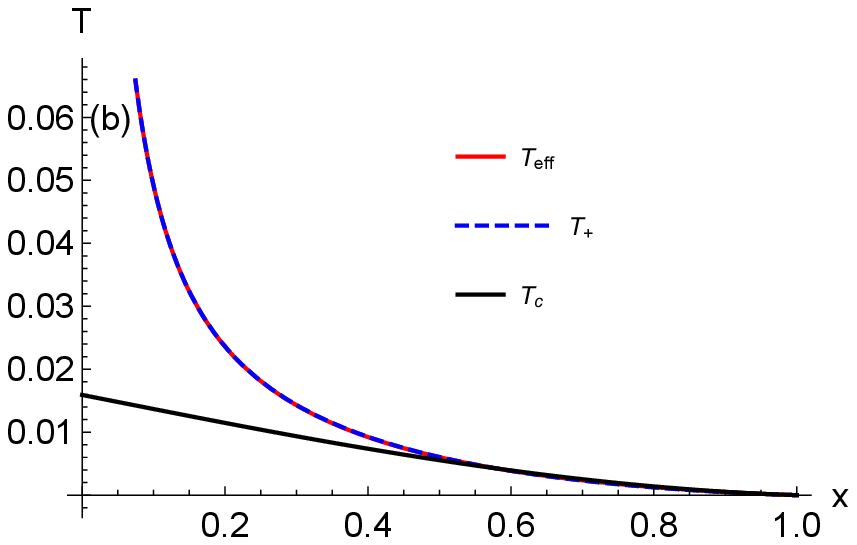}}}
\subfigure{{\includegraphics[width=0.3\columnwidth,height=1.5in]{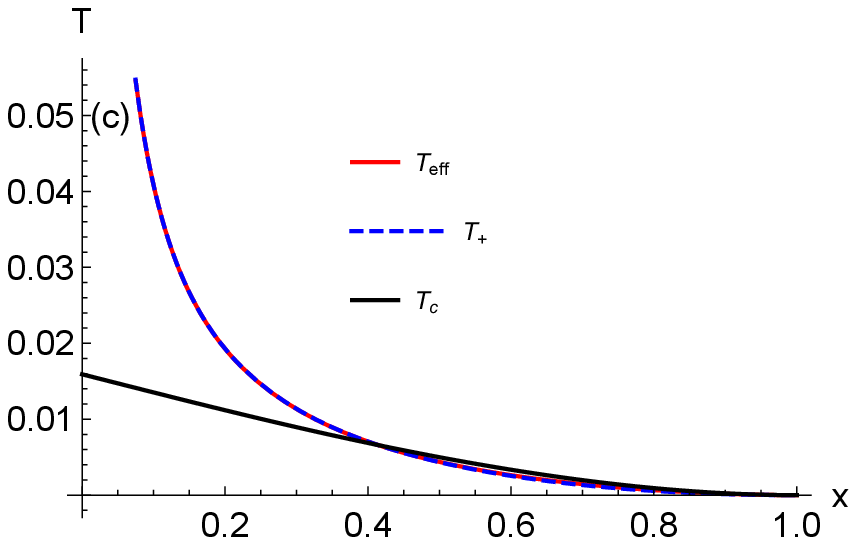}}}
\caption{(color online). ${T_{ + ,c,eff}} - x$ curves with ${l^2} = 100$, and (a) for $\phi_+^2= 0.3$, (b) for
$\phi_+^2= 0.4$, and (c) for $\phi_+^2 = 0.5$, respectively.}
\end{figure}
Substituting eq (2.14) into eqs (2.8), (2.11), and (2.12), all the temperatures on two horizons and the effective temperature can be rewritten as the function of $x$, $l$, and $\frac{Q^2}{r_c^2x^2}$. When fixed the parameters of $l$ and $\frac{Q^2}{r_c^2x^2}$, the graphs of temperature and effective pressure against the location ratio of the black hole horizon and the cosmological horizon with different electric potentials at the black hole horizon are given in Fig. 1 and Fig. 2. Note that for simplicity, we introduce the electric potential on black hole horizon as: $\phi_+\equiv\frac{Q}{r_c x}$.

From Fig. 1, when the cosmological constant ${l^{2}}$ is fixed, the relationship between the radiation temperature on the black hole horizon ${T_{+}}$ and the position ratio of two horizons. $x$ is the same as the one for the effective temperature ${T_{eff}}$. It indicates that the effective temperature ${T_{eff}}$ of the spacetime can be replaced by the radiation temperature on the black hole horizon ${T_{+}}$, when we study the relationship between the thermodynamic quantities and the position ratio of horizons $x$. Moreover, when the electric potential on black hole horizon increases, the position of the intersection point $x_{0}$ for the radiation temperature ${T_{+,c}}$ and the effective temperature ${T_{eff}}$ decreases. When the value of the electric potential on black hole horizon $\phi_{+}$ is small, the radiation temperature on cosmological horizon ${T_{c}}$ is less than the radiation temperature on the black hole horizon ${T_{+}}$. While when the value of $\phi_+^2$ is large, ${T_{+}}$ is less than ${T_{c}}$ in a certain interval of $x$. When $\phi_{+}^{2}>0.5$, both ${T_{+}}$ and ${T_{eff}}$ are negative, which does not meet the requirement of the equilibrium stability for thermodynamic system. Therefore, the upper bound
of $\phi_{+}^{2}$ is $0.5$.

\begin{figure}[tbp]
\centering
\subfigure{
{\includegraphics[width=0.3\columnwidth,height=1.5in]{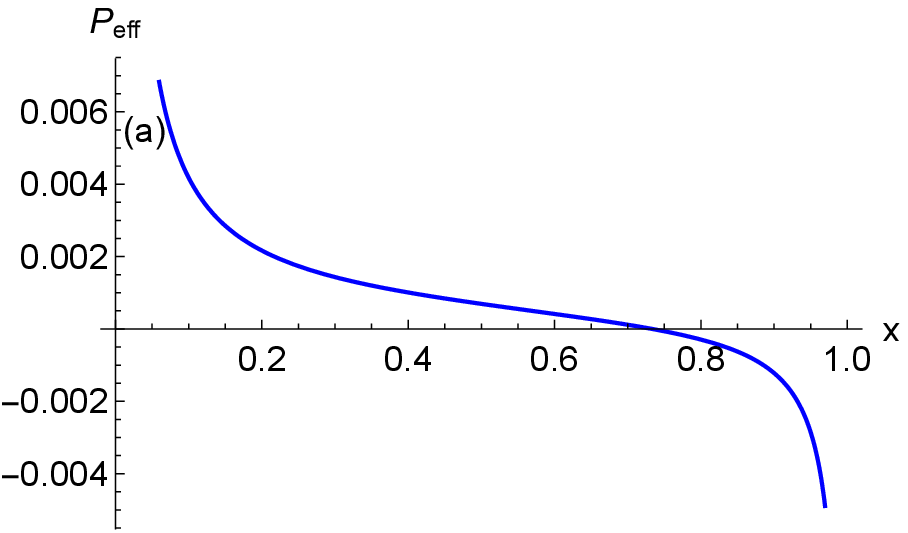}}
}
\subfigure{
{\includegraphics[width=0.3\columnwidth,height=1.5in]{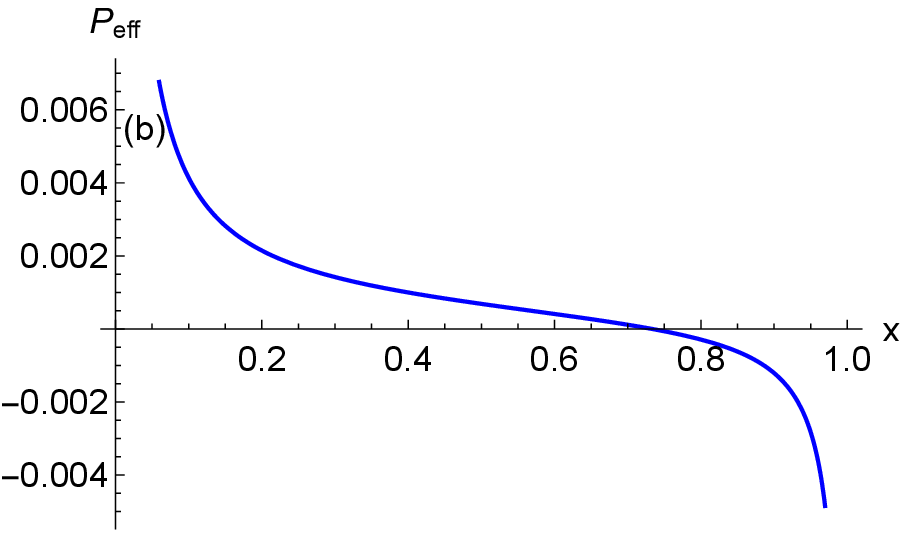}}
}
\subfigure{
{\includegraphics[width=0.3\columnwidth,height=1.5in]{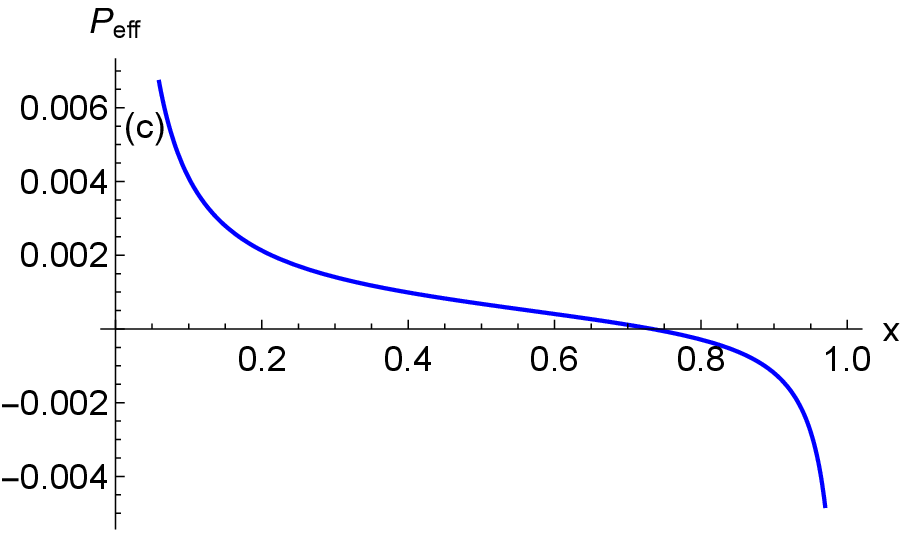}}
}
\caption{(color online). ${P_{eff}} - x$ curves with ${l^2} = 100$, and (a) for $\phi_+^2= 0.03$, (b) for
$\phi_+^2= 0.04$, and (c) for $\phi_+^2= 0.05$, respectively. }
\end{figure}

From Fig. 2, it is shown that when the cosmological constant is determined, the relationship between the effective pressure ${P_{eff}}$ and the position ratio $x$ is independent of the value of the electric potential on black hole horizon $\phi_+$. When $x$ approaches to 1, the effective pressure ${P_{eff}}$ is negative.

\section{Phase transition in RN-dS spacetime}

In the last section, the effective thermodynamic quantities of the RN-dS spacetime have been obtained. In this section, we will use the method similar to the study of the van der Waals system or the phase transition of the AdS black hole to investigate the thermodynamic property of the RN-dS spacetime. In the RN-dS spacetime, the heat capacities at constant volume and constant effective pressure are given as
\begin{eqnarray}
{C_V} &=& {T_{eff}}{\left( {\frac{{\partial S}}{{\partial {T_{eff}}}}}
\right)_V} = {T_{eff}}\frac{{{{\left( {\frac{{\partial S}}{{\partial {r_c}}}}
\right)}_x}{{\left( {\frac{{\partial V}}{{\partial x}}} \right)}_{{r_c}}} - {%
{\left( {\frac{{\partial S}}{{\partial x}}} \right)}_{{r_c}}}{{\left( {\frac{%
{\partial V}}{{\partial {r_c}}}} \right)}_x}}}{{{{\left( {\frac{{\partial V}%
}{{\partial x}}} \right)}_{{r_c}}}{{\left( {\frac{{\partial {T_{eff}}}}{{%
\partial {r_c}}}} \right)}_x} - {{\left( {\frac{{\partial V}}{{\partial {r_c}%
}}} \right)}_x}{{\left( {\frac{{\partial {T_{eff}}}}{{\partial x}}} \right)}%
_{{r_c}}}}},\\
{C_{{P_{eff}}}} &=& {T_{eff}}{\left( {\frac{{\partial S}}{{\partial {T_{eff}}}}%
} \right)_{{P_{eff}}}} = {T_{eff}}\frac{{\frac{{\partial S}}{{\partial x}}%
\frac{{\partial {P_{eff}}}}{{\partial {r_c}}} - \frac{{\partial S}}{{%
\partial {r_c}}}\frac{{\partial {P_{eff}}}}{{\partial x}}}}{{\frac{{\partial
{T_{eff}}}}{{\partial x}}\frac{{\partial {P_{eff}}}}{{\partial {r_c}}} -
\frac{{\partial {T_{eff}}}}{{\partial {r_c}}}\frac{{\partial {P_{eff}}}}{{%
\partial x}}}}.
\end{eqnarray}%
Then the ${C_V} - x$ and ${C_{peff}} - x$ curves with different parameters are plotted in Fig. 3 and Fig. 4.

\begin{figure}[tbp]
\centering
\subfigure{
{\includegraphics[width=0.3\columnwidth,height=1.5in]{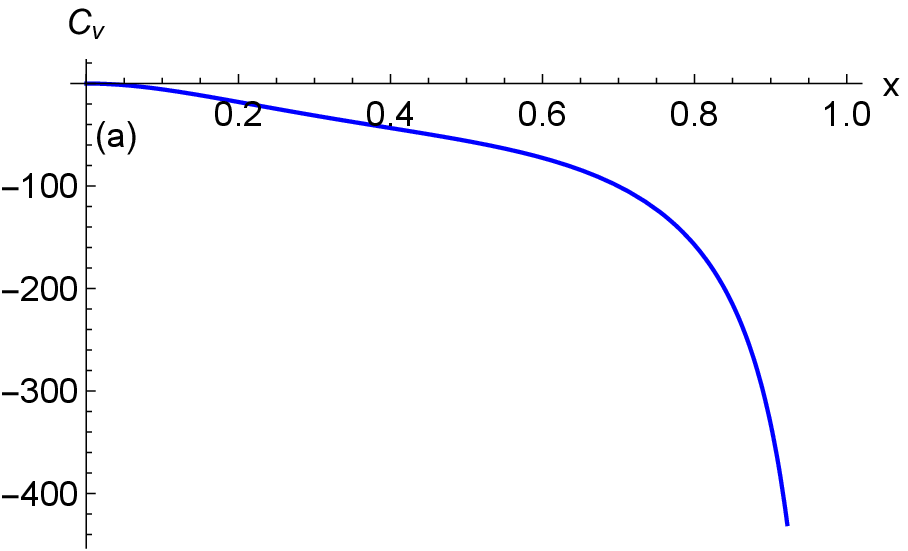}}
}
\subfigure{
{\includegraphics[width=0.3\columnwidth,height=1.5in]{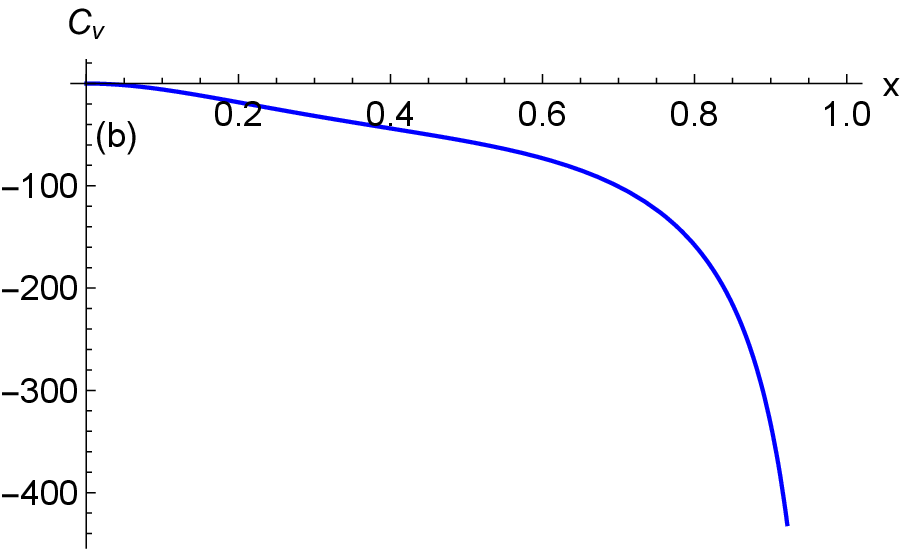}}
}
\subfigure{
{\includegraphics[width=0.3\columnwidth,height=1.5in]{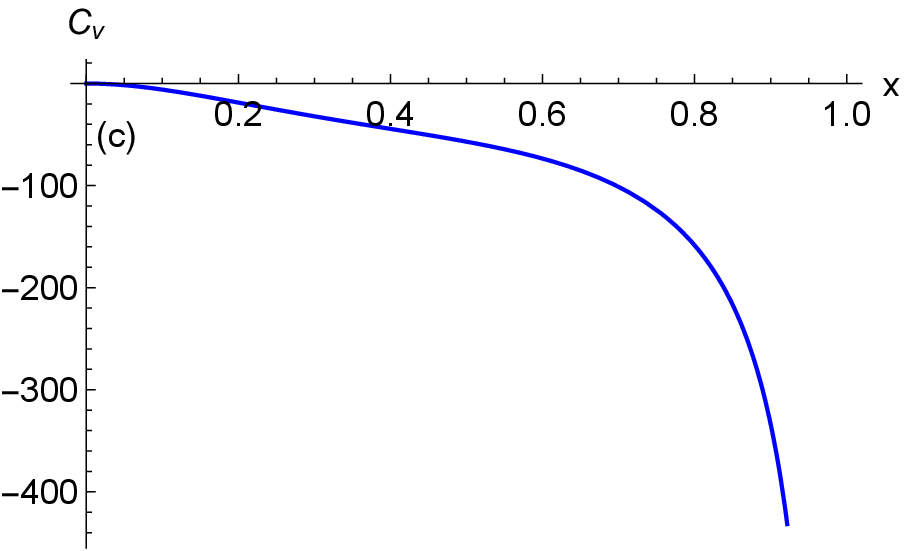}}
}
\caption{(color online). ${C_V} - x$ curves with ${l^2} = 100$, and (a) for $\phi_+^2= 0.03$, (b) for
$\phi_+^2= 0.04$, and (c) for $\phi_+^2= 0.05$, respectively. }
\end{figure}
\begin{figure}[tbp]
\centering
\subfigure{
{\includegraphics[width=0.3\columnwidth,height=1.5in]{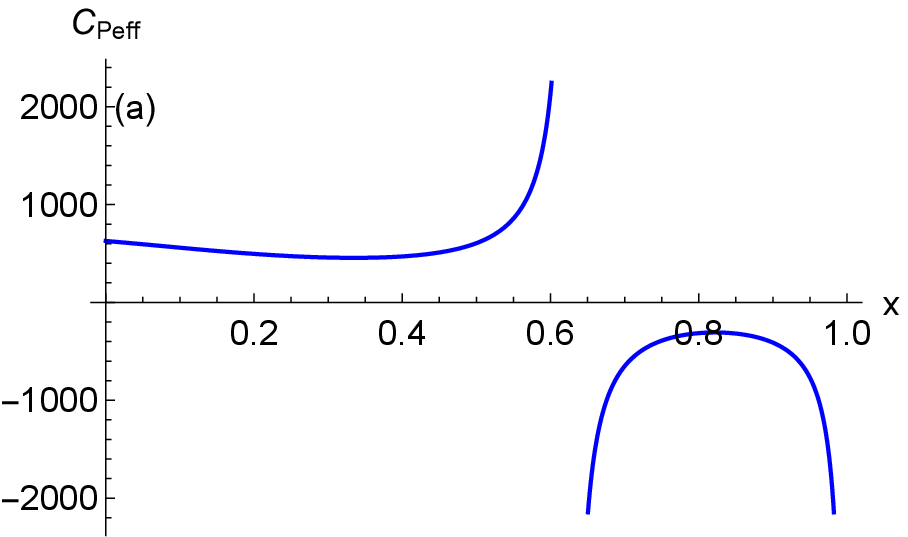}}
}
\subfigure{
{\includegraphics[width=0.3\columnwidth,height=1.5in]{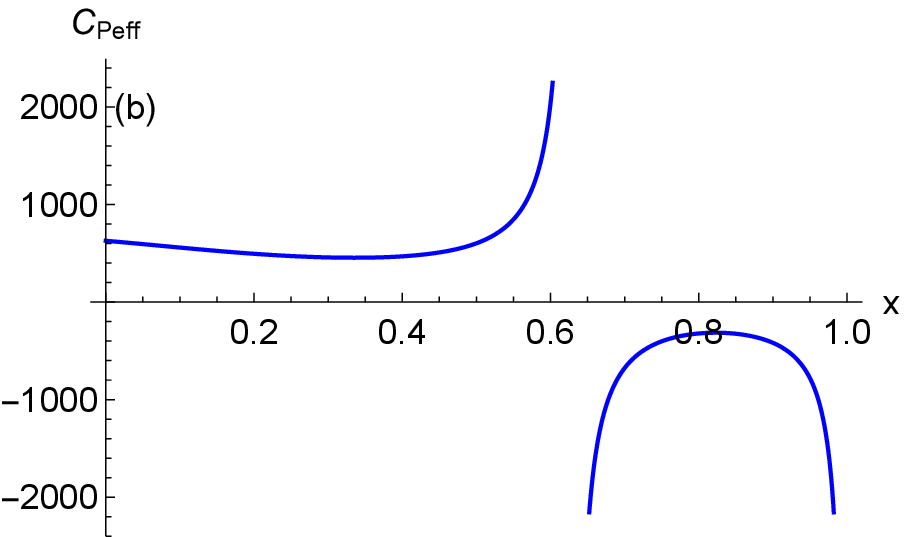}}
}
\subfigure{
{\includegraphics[width=0.3\columnwidth,height=1.5in]{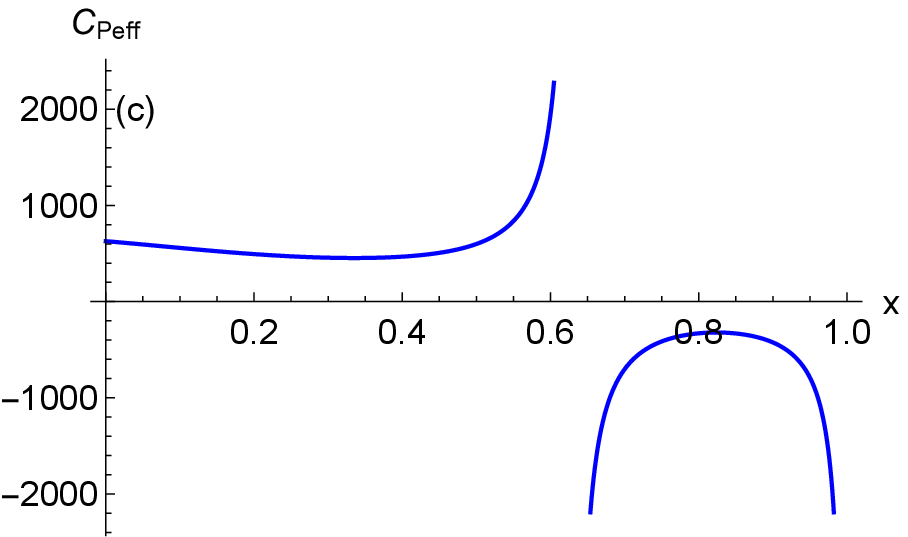}}
}
\caption{(color online). ${C_{peff}} - x$ curves with ${l^2} = 100$, and (a) for $\phi_+^2= 0.03$, (b) for
$\phi_+^2= 0.04$, and (c) for $\phi_+^2= 0.05$, respectively. }
\end{figure}

The volume expansion coefficient at a constant pressure is
\begin{equation}
\beta = \frac{1}{V}{\left( {\frac{{\partial V}}{{\partial {T_{eff}}}}}
\right)_{{P_{eff}}}} = \frac{1}{V}\frac{{\frac{{\partial V}}{{\partial x}}%
\frac{{\partial {P_{eff}}}}{{\partial {r_c}}} - \frac{{\partial V}}{{%
\partial {r_c}}}\frac{{\partial {P_{eff}}}}{{\partial x}}}}{{\frac{{\partial
{T_{eff}}}}{{\partial x}}\frac{{\partial {P_{eff}}}}{{\partial {r_c}}} -
\frac{{\partial {T_{eff}}}}{{\partial {r_c}}}\frac{{\partial {P_{eff}}}}{{%
\partial x}}}}.
\end{equation}
Then the $\beta - x$ curves with different parameters are plotted in Fig. 5.
\begin{figure}[tbp]
\centering
\subfigure{{\includegraphics[width=0.3\columnwidth,height=1.5in]{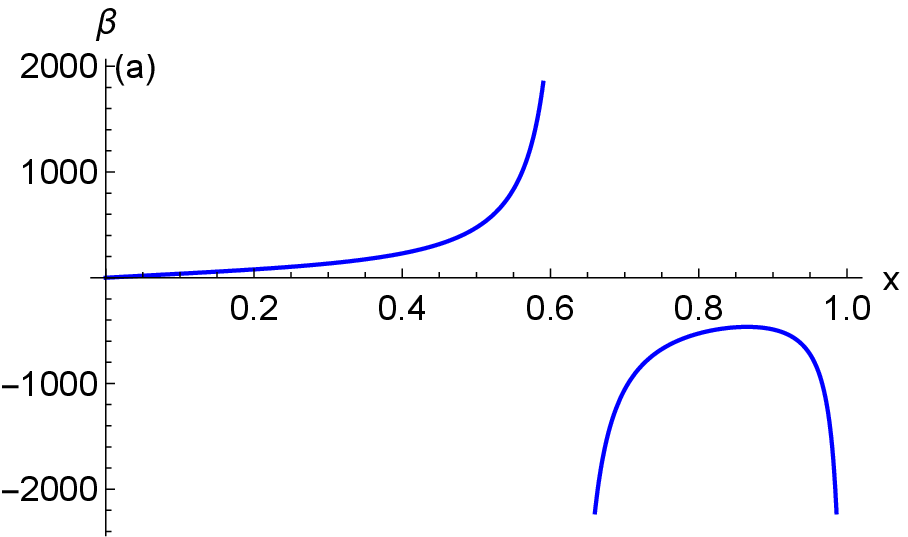}}}
\subfigure{{\includegraphics[width=0.3\columnwidth,height=1.5in]{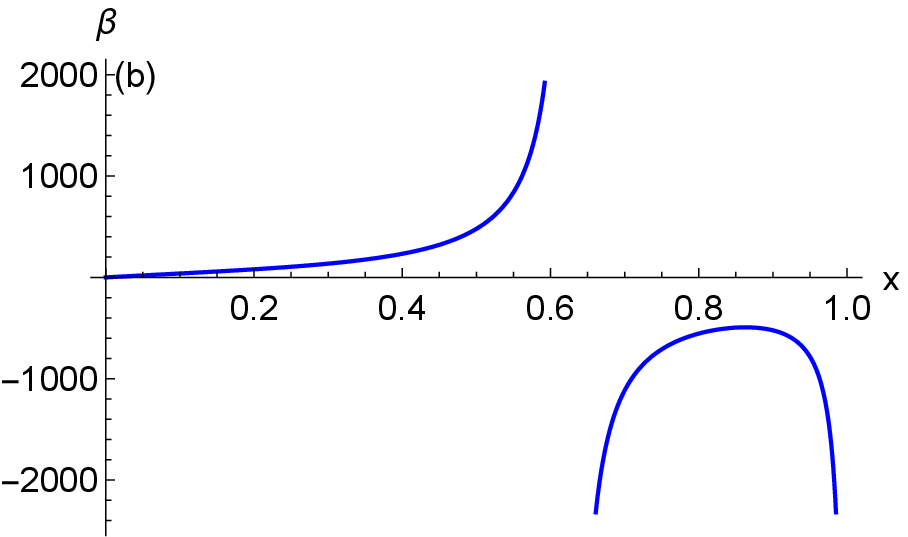}}}
\subfigure{{\includegraphics[width=0.3\columnwidth,height=1.5in]{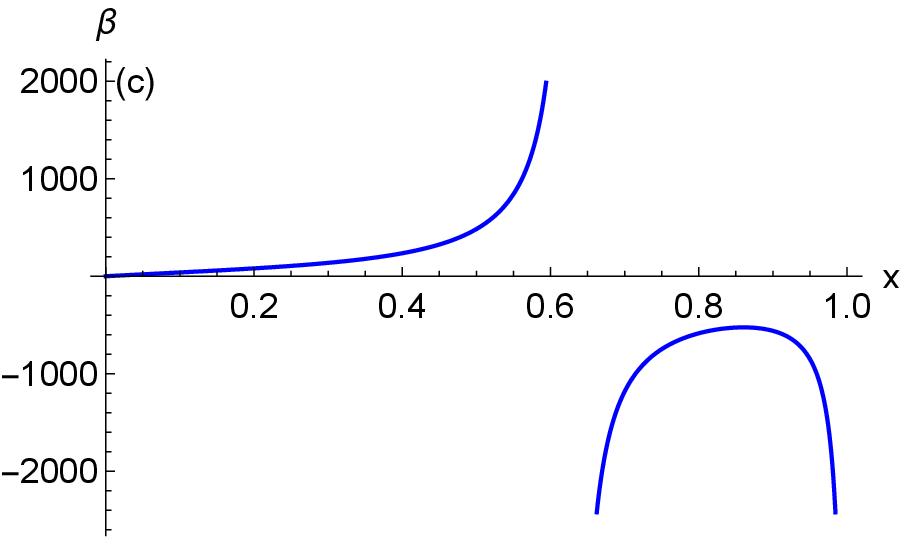}}}
\caption{(color online). $\beta  - x$ curves with ${l^2} = 100$, and (a) for $\phi_+^2= 0.03$, (b) for
$\phi_+^2= 0.04$, and (c) for $\phi_+^2= 0.05$, respectively. }
\end{figure}
The isothermal compression factor is
\begin{equation}
{\kappa _{{T_{eff}}}} = - \frac{1}{V}{\left( {\frac{{\partial V}}{{\partial {%
P_{eff}}}}} \right)_{{T_{eff}}}} = \frac{1}{V}\frac{{\frac{{\partial V}}{{%
\partial x}}\frac{{\partial {T_{eff}}}}{{\partial {r_c}}} - \frac{{\partial V%
}}{{\partial {r_c}}}\frac{{\partial {T_{eff}}}}{{\partial x}}}}{{\frac{{%
\partial {T_{eff}}}}{{\partial x}}\frac{{\partial {P_{eff}}}}{{\partial {r_c}%
}} - \frac{{\partial {T_{eff}}}}{{\partial {r_c}}}\frac{{\partial {P_{eff}}}%
}{{\partial x}}}}.
\end{equation}
Then the ${\kappa _{{T_{eff}}}} - x$ curves with different parameters are plotted in Fig. 6.

\begin{figure}[tbp]
\centering
\subfigure{
{\includegraphics[width=0.3\columnwidth,height=1.5in]{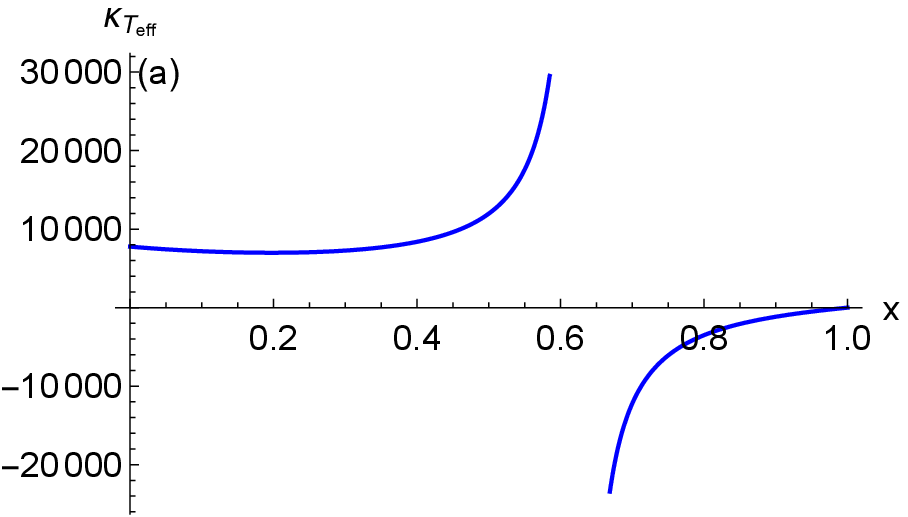}}
}
\subfigure{
{\includegraphics[width=0.3\columnwidth,height=1.5in]{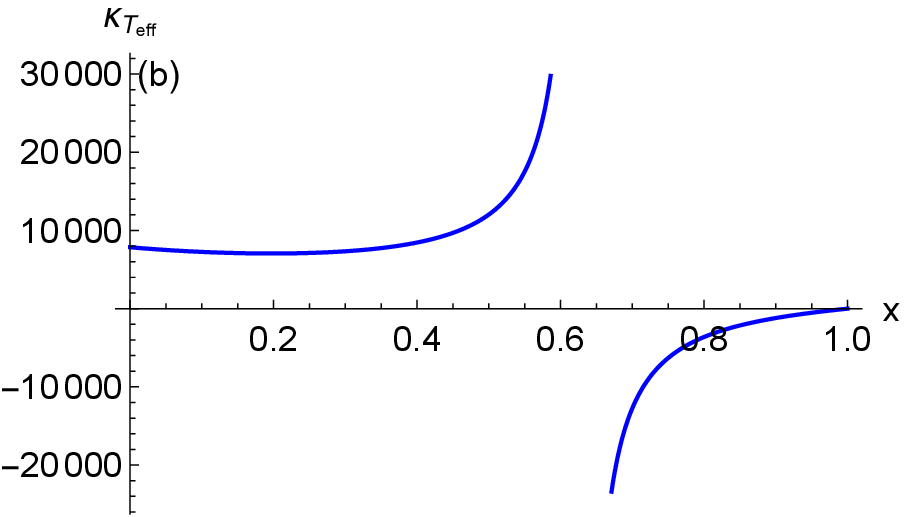}}
}
\subfigure{
{\includegraphics[width=0.3\columnwidth,height=1.5in]{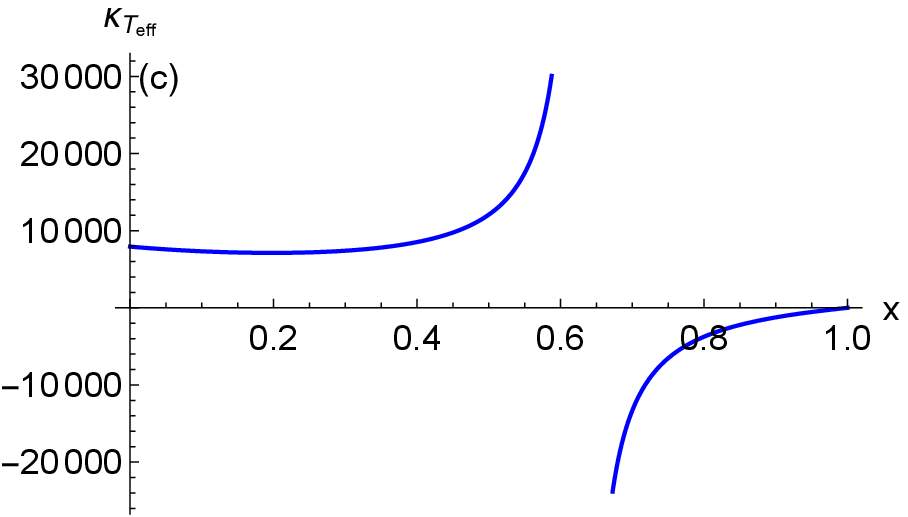}}
}
\caption{(color online). ${\kappa _{{T_{eff}}}} - x$ curves with ${l^2} = 100$, and (a) for $\phi_+^2= 0.03$, (b) for
$\phi_+^2= 0.04$, and (c) for $\phi_+^2= 0.05$, respectively. }
\end{figure}

In RN-dS spacetime, Helmholtz free energy can be written as $F = M - {T_{eff}}S$. From Fig. 3, it is known that the heat capacity at a constant volume is not zero, i.e., ${C_V} \ne 0$. This is different from the case in AdS spacetime, but similar to the case of Schwarzschild black hole within $0 < x \le 1$. According to the requirement of the equilibrium stability for ordinary thermodynamic system, the RN-dS spacetime with a constant volume is an unstable thermodynamic system.

\begin{figure}[tbp]
\centering
\subfigure{
{\includegraphics[width=0.3\columnwidth,height=1.5in]{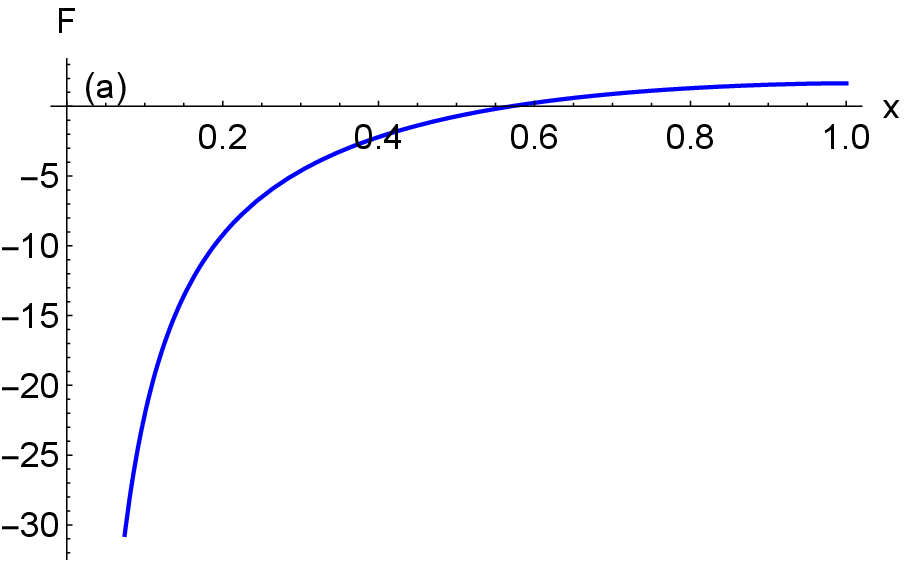}}
}
\subfigure{
{\includegraphics[width=0.3\columnwidth,height=1.5in]{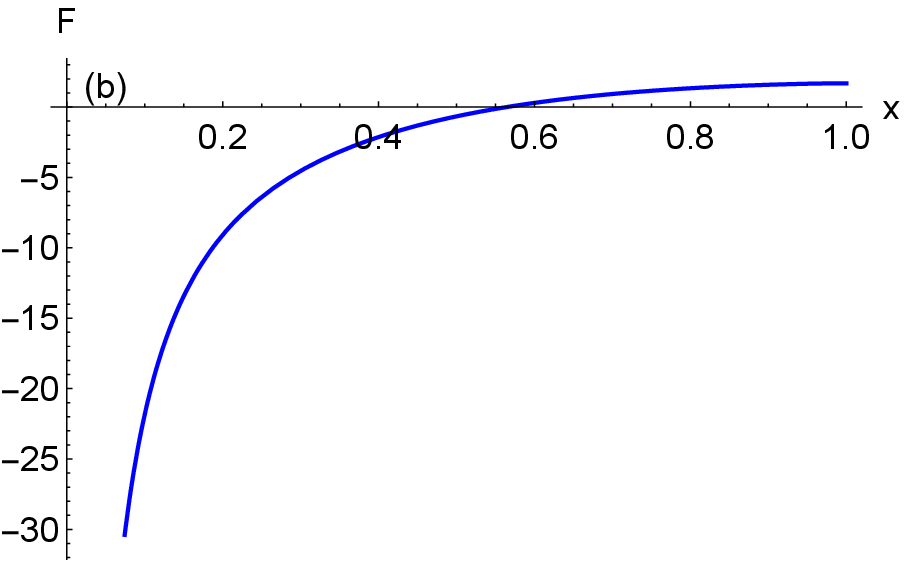}}
}
\subfigure{
{\includegraphics[width=0.3\columnwidth,height=1.5in]{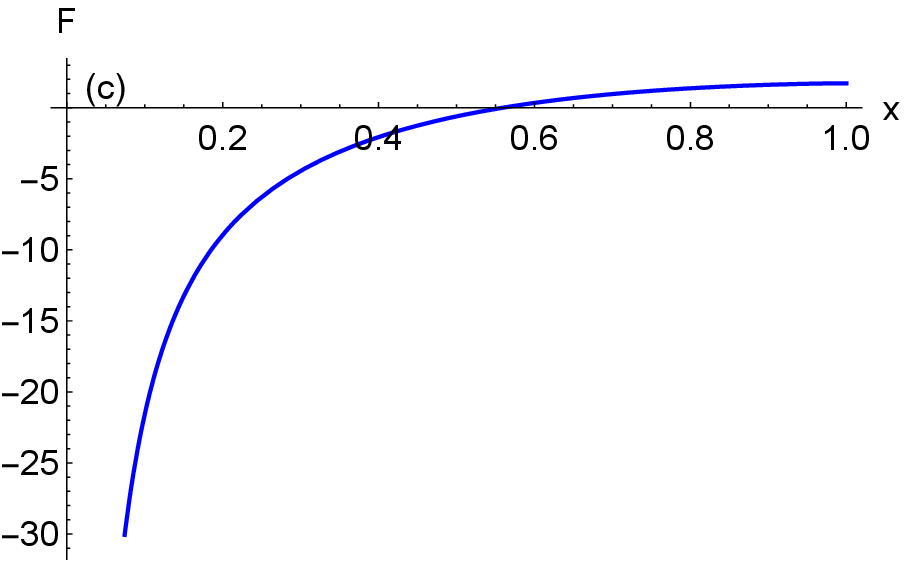}}
}
\caption{(color online). $F - x$ curves with ${l^2} = 100$, and (a) for $\phi_+^2= 0.03$, (b) for $\phi_+^2=
0.04$, and (c) for $\phi_+^2= 0.05$, respectively. }
\end{figure}


From Fig. 1 and Fig. 2, it is shown that when the cosmological constant $l^{2}$ and the electric potential on the black hole horizon $\phi_ + $ are determined, the effective temperature $T_{eff}$ and the pressure $P_{eff}$ of the spacetime are monotonic functions of the position ratio $x$. Therefore, the results given in Fig. 7 are the same, i.e., the curves $G - x$, ${{{T_{eff}}}} - x$ and ${{{P_{eff}}}} - x$ are the same. Since the Helmholtz function $F$ of spacetime is a single valued function for the position ratio $x$ , there is no first-order phase transition in RN-dS spacetime. From Fig. 4 to Fig. 6, we can find that RN-dS spacetime has a second-order phase transition similar to that of van der Waals system.

\section{Entropic force between two horizons}

The authors in Ref. \cite{30} had made a guess of between gravity and entropic force. That was later proved
\cite{31,32,33,34,35,36,37} in a classical scenario. In addition people proposed that gravity is a manifestation of material information through combining a thermal gravitation treatment to Hooft's holographic principle. Accordingly, gravitation
ought to be viewed as an emergent phenomenon. Such exciting Verlinde's idea received a lot of attention \cite{38,39,40,41,42,43,44,45}.

In this section, the interaction force between two horizons will be studied in the circumstances that the cosmological constant ${l^2}$ is a constant. To extend the expression of the entropic force to the interaction force between two horizons in RN-dS spacetime, the formula of the entropic force is
\begin{equation}
\tilde F(x) = {T_{eff}}\frac{{{{\left( {\frac{{\partial \tilde S}}{{\partial
{r_c}}}} \right)}_x}{{\left( {\frac{{\partial {T_{eff}}}}{{\partial x}}}
\right)}_{{r_c}}} - {{\left( {\frac{{\partial \tilde S}}{{\partial x}}}
\right)}_{{r_c}}}{{\left( {\frac{{\partial {T_{eff}}}}{{\partial {r_c}}}}
\right)}_x}}}{{(1 - x){{\left( {\frac{{\partial {T_{eff}}}}{{\partial x}}}
\right)}_{{r_c}}} + {r_c}{{\left( {\frac{{\partial {T_{eff}}}}{{\partial {r_c%
}}}} \right)}_x}}},
\end{equation}
where
\begin{equation}
\tilde S = \pi r_c^2f_0(x) = \pi r_c^2\left( {\frac{8}{5}{{\left( {1 - {x^3}}
\right)}^{2/3}} - \frac{{2\left( {4 - 5{x^3} - {x^5}} \right)}}{{5\left( {1
- {x^3}} \right)}}} \right).
\end{equation}
From Eq. (2.7), it is known that $\tilde S$ is the entropy arisen from the interaction between the two horizons.

\begin{figure}[tbp]
\centering

\includegraphics[width=0.5\columnwidth,height=1.5in]{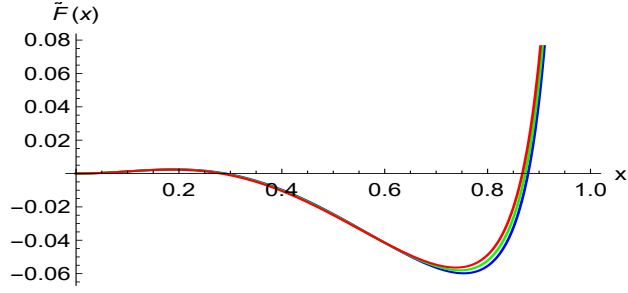}

\caption{(color online). $\tilde F(x) - x$ curves with $\frac{1}{{r_c^2}} = \frac{{(1 + x + {x^2})}}{{{l^2}(1 - x{Q^2}/r_ + ^2)}}$, and
solid blue line for $\phi_+^2= 0.03$,  solid green line for $\phi_+^2= 0.04$, and  solid red line for
$\phi_+^2= 0.05$, respectively. }

\end{figure}

It can be seen from Fig. 8, when the cosmological constant $l^{2}$ is determined, and horizons of black holes and the universe tend to coincide, i.e.,$\mathrm{x} \to 1$, the entropic force between the two horizons tends to infinity. Without any other forces, the two horizons are separated from each other by the entropic force between them. With the separation of the two horizons, i.e., the value of $x$ decreases, the entropic force between the two horizons turns to attraction from repulsion. Finally, the spacetime tends to pure dS spacetime, and the entropic force tends to zero.

Comparing the $\tilde F(x) - x$ curve given in Fig. 8 with the curve of the Lennard-Jones force changed with particles positions \cite{46,47}, we find that the results obtained by completely different methods are very similar. This indicates that the Lennard-Jones force between two particles has a certain internal relationship with the entropic force between the two horizons. It provides indirect experimental data to further explore the entropic force between the two horizons.

\section{Conclusions}

Black hole physics, especially black hole thermodynamics, is directly related to various fields of physics, such as gravitation, statistical physics, theories of particle and field. This makes it receive much more attention. It can be said that black hole physics has become a test field for relevant theories, and makes different physical subjects connect to each other profoundly and fundamentally, where black hole thermodynamics plays an important role. For a de Sitter spacetime, since the radiation temperatures on the black hole horizon and the cosmological horizon is generally different, the spacetime does not meet the requirement of the equilibrium stability, which makes some difficulties in the study of the thermodynamic characteristics of de Sitter spacetime. Recently, by assuming that the entropy of de Sitter spacetime is the sum of the entropy on the two
horizons \cite{5,12,13,17,24}, the thermodynamic characteristics of de Sitter spacetime has been studied. It shows that when the radiation temperature on black hole horizon is equal to the radiation temperature on the cosmological horizon, i.e., $T=T_{+}=T_{c}$, the effective temperature is not equal to the radiation temperature on the above mentioned two horizons, i.e., $T_{eff} \ne T$. This result is hard to accept. Moreover, there is no theoretical proof for the assumed entropy when the effective temperature of de Sitter spacetime is studied.

In this paper, using conditions that the thermodynamic quantities of de Sitter spacetime satisfy the first law of thermodynamics, and when $T = {T_ + } = {T_c}$, $T = {T_{eff}}$, the entropy and the effective thermodynamic quantities of de Sitter spacetime have been obtained. From the effective thermodynamic quantities, it is found that when the cosmological constant $l^{2}$ is fixed and the position ratio $x$ of the two horizons takes a certain value, a second-order phase transition can happen in the spacetime. When the heat capacity at constant volume is negative, i.e., ${C_V} < 0$, the spacetime is unstable. Without considering other forces, horizons of black holes and the universe are separated or closed to each other by the entropic force between them.

From modern cosmology, it is known that if the universe is determined by dark energy, it will expand forever. While on the opposite side, the expansion of our universe will be stopped and then turned to contraction. In this work, we find that the entropic force plays a certain role in the cosmic expansion and contraction. Different matter fields in the universe
make different effects on the cosmic expansion. In de Sitter spacetime, the result of the entropic force between horizons changed with the position ratio shows that when the value of the electric potential on black hole horizon $\phi_ +$ increases, the range of the repulsive force between the two horizons also increases. It indicates that the electric potential on black hole horizon in de Sitter spacetime affects the cosmic expansion.

Within the framework of general relativity, the entropic force arisen from the interaction between the black hole horizon and the cosmological horizon is very similar to the Lennard-Jones force between two particles confirmed by experiments. Our results indicate the relationship between quantum mechanics and thermodynamics. This provides a new way to research not only the state of particles and the interaction between them in black holes, but also the Lennard-Jones potential and the state of particles in
ordinary thermodynamic systems.

\section*{Acknowledgements}

We thank Prof. Z. H. Zhu for useful discussions.

This work was supported by the National Natural Science Foundation of China (Grant Nos. 12075143, 11971277, 11705106, 11705107, 11605107),
the Scientific and Technological Innovation Programs of Higher Education Institutions of Shanxi Province, China (Grant No. 2020L0471, No.
2020L0472, No. 2019L0743) and the Science Technology Plan Project of Datong City, China (Grant No. 2020153).

\end{document}